\documentclass[%
 reprint,
 amsmath,amssymb,
 aps,
]{revtex4-2}

\usepackage{graphicx}
\usepackage{soul}
\usepackage{dcolumn}
\usepackage{bm}
\usepackage{color}
\usepackage{soul}
\usepackage{eurosym}

\begin{document}
\preprint{APS/123-QED}
\title{Intensity-Correlation Synthetic Wavelength Imaging in Dynamic Scattering Media}

\author{Khaled Kassem$^1$, Areeba Fatima$^1$, Patrick Cornwall$^2$, Muralidhar Madabhushi Balaji$^2$, Daniele Faccio$^{1,2,*}$, Florian Willomitzer$^{2,}$ }
\email{daniele.faccio@glasgow.ac.uk; fwillomitzer@arizona.edu}
\affiliation{$^1$ School of Physics and Astronomy, University of Glasgow, United Kingdom\\
$^2$ Wyant College of Optical Sciences, University of Arizona, Tucson, AZ, USA 85721\\}


\begin{abstract} 
Imaging through dynamic scattering media, such as biological tissue, presents a fundamental challenge due to light scattering and the formation of speckle patterns. These patterns not only degrade image quality but also decorrelate rapidly, limiting the effectiveness of conventional approaches, such as those based on transmission matrix measurements. 
Here, we introduce an imaging approach based on second-order correlations and synthetic wavelength holography (SWH) to enable robust image reconstruction through thick and dynamic scattering media. By exploiting intensity speckle correlations and using short-exposure intensity images, our method computationally reconstructs images from a hologram without requiring phase stability or static speckles, making it inherently resilient to phase noise. 
Experimental results demonstrate high-resolution imaging in both static and dynamic scattering scenarios, offering a promising solution for biomedical imaging, remote sensing, and real-time imaging in complex environments.
\end{abstract}
\maketitle

\begin{figure*}[t!]
 \centering
 \includegraphics[width=1\textwidth]{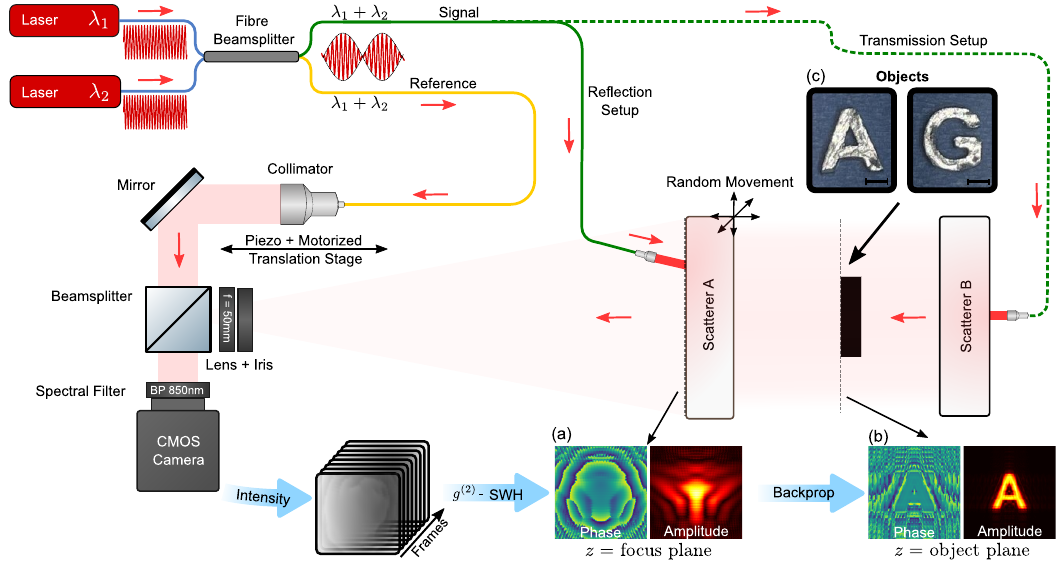}
\caption{\textbf{Experimental setup overview.}  
Overview of the measurement process. The two lasers are combined at a fiber beamsplitter, ensuring that both wavelengths, $\lambda_1$ and $\lambda_2$, are present in both the signal and reference arms. Two measurement configurations are depicted: reflection (solid green) and transmission (dashed green). The reference arm collimator is adjusted using a motorized translation stage and rapidly moved with a piezo stage for phase randomization (PR). 
(a) SWL hologram (simulation) generated from 3 sets of intensity images at the surface of the scatterer.  
(b) Hologram after numerical backpropagation to the hidden object plane, revealing the object.  
(c) Photographs of the sample objects (3D-printed letters) with a scale bar of 5 mm.}
\label{fig:experiment}
\end{figure*}

\textbf{Introduction.}
Imaging through scattering media—such as fog, biological tissue, or smoke—is essential for applications including medical diagnostics, autonomous vehicle navigation, and remote sensing in low-visibility or turbulent environments.
\cite{soldevila2023functional,yoon2022recent,bijelic2020seeing,locatelli2013imaging}.\\ However, multiple scattering in these media leads to chaotic speckle patterns, rendering traditional imaging methods (based e.g. on just a camera) ineffective in reconstructing the object.
Various techniques have been developed to tackle the challenge of imaging in \emph{static} complex media, including wavefront shaping \cite{Feng2023NeuWS:Media,Gigan2022RoadmapMedia,Sanjeev2019Non-InvasiveTechnique,Nixon2013Real-timeFeedback,Yu2015RecentApplications}, transmission matrix (TM) methods \cite{yaqoob2008,popoff2010,vellekoop2007,Blochet2017Transmission-matrix-basedMedium}, and speckle-correlation techniques \cite{Katz2014Non-invasiveCorrelations,Jauregui-Sanchez2022TrackingCorrelations,bertolotti2012}. While these approaches have shown significant success, they provide limited performance through thick scatterers. More importantly, their effectiveness is often compromised in dynamic environments where the slightest movements lead to a complete decorrelation of the speckle patterns.\\
To counteract these affects, speckle deblurring techniques have been explored, but they remain limited by speckle coherence and typically work only for simple objects \cite{Zhang2023High-throughputDe-blurring}. Machine learning (ML)-based approaches have also been investigated, leveraging data-driven models to reconstruct images \cite{Sun2019ImageLearning,Starshynov2022StatisticalMedia,ChopiteDeepReconstruction,Liu2024Learning-basedMedia}. However, these methods often require extensive labelled datasets and lack physical priors, limiting their adaptability to highly dynamic scenarios where e.g. both microscopic and macroscopic properties change in time. Quantum imaging techniques utilizing entangled photon pairs offer potential solutions for imaging through scattering \cite{defienne2021polarization}, yet their reliance on polarization entanglement makes them unsuitable for strongly scattering media where the polarization is changed.\\
In this context, synthetic wavelength holography (SWH) \cite{Willomitzer2021FastHolography,willomitzer2024synthetic} has emerged as a relatively recent technique for imaging through scattering media that exploits speckle correlations to mitigate the effect of scattering. Here, synthetic waves resulting from the beating between two coherent light waves at closely placed wavelengths traverse through the scattering media. The synthetic field assembled 
 at the far end of the scattering medium has been shown to be the holographic representation of the hidden object placed in the media. These holograms preserve the phase information at scales exceeding the synthetic wavelengths, hence overcoming the deleterious effects of scattering.  \\
However, this approach requires capturing complex-valued fields at closely spaced optical wavelengths for hologram generation, necessitating the use of holographic acquisition methods. Although it has been shown that this process can be performed in single-shot \cite{Ballester2024Single-shotSensors,cornwall2024single}, optical phase stability during the camera exposure is still required, which imposes space-time-bandwidth restrictions on current acquisition systems.\\
In this work, we introduce a paradigm that relaxes these constraints by exploiting the second-order correlation i.e., the $g^{(2)}$  correlations. As these $g^{(2)}$  correlations can be evaluated from \emph{intensity} only measurements at different time instances, the need for holographic acquisition apparatus is no longer needed and the robustness of the system to phase fluctuations is significantly improved.
In general, the $g^{(2)}$ correlations have been studied for high-resolution, phase-robust microscopy in both quantum applications \cite{ndagano2022quantum} and classical approaches inspired by quantum sensing techniques, such as Hong-Ou-Mandel (HOM) interference with two beating wavelengths \cite{Johnson2023TowardInterferometry, kassem2025time}.
While intensity correlations based interferometers are known to be immune to time-dependent phase fluctuations, they  still need spatial coherence, which is often absent in dynamic scattering environments \cite{Szuniewicz2023NoiseCorrelation,Thekkadath2023IntensityLight}. Furthermore, other correlation-based holography techniques \cite{Luo2024Single-ShotHolography} demand quick measurements to track rapidly evolving speckle patterns, a challenge with highly scattering and low SNR conditions. Spatial speckle correlations at two closely spaced wavelengths have been shown to extract information from scattering media \cite{bett2024illumination}. However, the approach works best for sparse objects to get significant correlation signals, and requires large spatial sampling.\\
In our new approach, we reformulate the SWH principles under the framework of $g^{(2)}$  correlation formalism. We show that the mathematical implication developed therein lends itself well to the application of imaging through a dynamic scattering media such that the inherent fluctuation of the dynamic media can be harnessed for image reconstruction. The efficacy of our approach is shown across various scenarios, including traditional static two-planes of scatter settings, complex volume scattering, and dynamic scattering environments.
To further enhance image reconstruction, we integrate a total variation (TV)-based denoising algorithm and iterative backpropagation. By combining both physical and computational approaches, our method not only provides a robust solution for imaging through dynamic scattering media but also shows how dynamic scattering can be utilized as an asset rather than a limitation.\\
\textbf{Theory.}
In conventional interferometry (e.g., in a Mach-Zehnder interferometer), if $E_i$ ($i$ = 1,2) are two electric fields in the two arms of the interferometer, then electric fields at one of the two output ports denoted by \(E\) can be expressed as:
\begin{equation}\label{eq:E_out_port_A}
    E(t_1,t_2,t) = \frac{1}{\sqrt{2}} \left[ E_1(t - t_1) + j E_2(t - t_2) \right]
\end{equation}  
where \(t\) is the global time, and \(t_i\) represents the time taken by the electric field $E_i$ to traverse the path of the interferometer. Assuming plane waves with equal amplitudes for the electric fields in both paths, we express the individual fields as:

\begin{equation}\label{eq:E_field}
    E_i(t) = |E_0| e^{j(\omega_i t + \varphi_i)},
\end{equation}

where \(|E_0|\) denotes the amplitude of the electric field, \(\omega_i\) is the frequency of the field, and \(\varphi_i\) is the phase. 
The measured intensity then at the output port, can be defined as:
\begin{equation}\label{eq:int_e_field}
    I(t_1,t_2) = \langle E^*(t_1,t_2,t) E(t_1,t_2,t) \rangle,
\end{equation}
For the case when the electric fields in the arms of the interferometer have same frequency ($\omega_1 = \omega_2 = \omega$) and the spatial phase difference between the two is $\Delta \varphi$ ($= \varphi_1-\varphi_2$) this intensity expression can be evaluated using Eq.\ref{eq:E_out_port_A} and Eq.\ref{eq:E_field} in Eq.\ref{eq:int_e_field} to give the following (see SM):
\begin{equation}\label{int_at_port}
    I(\Delta t) = |E_0|^2 \left( 1 + |g^{(1)}_{(\Delta t)}| \sin\left(\omega \Delta t - \Delta \varphi\right) \right).
\end{equation}
where the modulation amplitude $g^{(1)}$ is the first-order correlation function, also called the field correlation function, and is defined as \cite{loudon2000quantum}:
\begin{equation}\label{eq:g1general}
    g^{(1)}(\tau=0, \Delta t) = \frac{\langle E_1^*(t, \Delta t) E_2(t, \Delta t)\rangle_t}{\sqrt{\langle |E_1(t, \Delta t)|^2 \rangle \langle |E_2(t, \Delta t)|^2 \rangle}},
\end{equation}
Here, $\tau$ denotes the relative time difference between the two electric fields after both fields have arrived at the output port (and therefore different from $\Delta t$, which has been accounted for before arriving at the port). We only consider the case when $\tau$ is zero. Since we assume monochromatic light, this implies that $g^{(1)}$ remains constant for all $\Delta t$.
The absolute value of $g^{(1)}$ can be interpreted as the visibility of the interference fringes, varying between 0 (uncorrelated) and 1 (fully correlated) \cite{loudon2000quantum}.\\
In this framework, the field for a wavelength $\lambda_m$ can be expressed as:
\begin{equation}\label{eq:4}
E(\lambda_m) = \left|g^{(1)}_{\lambda_m}\right| e^{i \varphi(\lambda_m)}
\end{equation}
We leverage these formulations in the field of SWI. Willomitzer \textit{et al}. \cite{Willomitzer2021FastHolography} defined the computational construction of the synthetic field  as the product of the electric field at wavelength $\lambda_1$ and the complex conjugate of the electric field at wavelength $\lambda_2$. Reformulating this definition in terms of the field correlation function using Eq.(\ref{eq:4}), we have (valid up to a proportionality constant):
\begin{equation}\label{eq:synfield}
    E(\Lambda) = 
    \left|g^{(1)}_{\lambda_1}\right| \left|g^{(1)}_{\lambda_2}\right| e^{i \Phi(\Lambda)} 
\end{equation}
where we have introduced the beat wavelength, $\Lambda=\frac{\lambda_1\lambda_2}{\lambda_1-\lambda_2}$, and the phase $\Phi(\Lambda) = \phi_1(\lambda_1)-\phi_2(\lambda_2)$.\\ This expression relates the synthetic field to the first-order coherence function, which, in turn, is related to the electric field at the optical wavelengths. This implies that, in this framework, constructing the synthetic field requires measuring the amplitude of each of the optical electric fields, which can be cumbersome. To this end, previous work based on holographic methods, such as a lock-in camera approach, off-axis holography, or conventional digital holography {\cite{Willomitzer2021FastHolography,Ballester2024Single-shotSensors,Kotwal2022Swept-AngleInterferometry,cornwall2024single}, have typically been employed to generate the synthetic field}. These methods generally involve separately measuring the hologram for each wavelength and subsequently generating the synthetic wavelength hologram computationally as mentioned above. Although single-shot variants of this procedure have been demonstrated that provide robustness to dynamics to a certain extent \cite{Ballester2024Single-shotSensors,cornwall2024single}, holographic methods still depend on the local phase $\varphi$ of the optical wave, which makes them highly susceptible to phase noise.\\
In the following, we show that this dependence can be eliminated by utilizing the second-order correlations instead that can be agnostic to such phase noise. The second-order correlation function is defined as  
\begin{equation}\label{eq:g2generalintens}
    g^{(2)}(\tau, \Delta t) = \frac{\langle I(t, \Delta t) I(t + \tau, \Delta t)\rangle_t}{\langle I(t, \Delta t) \rangle_t^2},
\end{equation}  
where $\langle \cdot \rangle_t$ denotes the ensemble average. The $g^{(2)}$ defined above can be evaluated by using the intensity expression of Eq.\ref{int_at_port}.\\ So far the analysis presented here deals with the case when the electric fields entering the interferometer have a single wavelength band of non-zero bandwidth, resulting in matched frequencies in the two arms of the interferometer. We can extend this analysis for the case where two distinct fields with different frequencies enter the interferometer. The $g^{(2)}$ can be calculated as a cross-correlation of the intensity measured for one wavelength with the intensity measured for another wavelength. 
This would require to separate the intensity of both wavelengths, e.g., by spectrally filtering at the two output ports of a Mach-Zehnder interferometer. It can be shown that the second-order cross-correlation obtained in this case is given by (see Supplementary Material for details)

\begin{equation}\label{g(2)_explicit}
\begin{aligned}
    g^{(2)}_{x}(\Delta t) = \langle 1 - 0.5|g^{(1)}_{(\Delta t,I)}||g^{(1)}_{(\Delta t,II)}|\\ \left[\cos\left(\Delta \omega \Delta t \right) - \cos\left((\omega_{1} + \omega_{2}) \Delta t - 2\Delta \varphi\right) \right] \rangle.
\end{aligned}
\end{equation}

Here, $\omega_1$ and $\omega_2$ are the frequencies of the electric field entering the interferometer. A key point here is that to properly obtain the time-averaged quantities such that the statistical properties of the fields are correctly modeled, phase randomization over different ensembles is required. Thus, the ensemble average in Eq.~\ref{eq:g2generalintens} can be considered as a phase-randomized time average. This forms the crux of our proposed method as it implies that we must actively vary the phase during the ensemble averaging to accurately measure \( g^{(2)} \). As discussed later, we show that this requirement can be leveraged to image through dynamic scattering media. Now, the second cosine term in Eq.~\ref{g(2)_explicit} contains the random phase term ($\Delta\varphi$) and will therefore average to zero due to the ensemble averaging.\\
So far, we have discussed the case where no object is placed in the arms of the interferometer. In the case where there is an object, both fields with different wavelengths entering the interferometer will accumulate their own phases ($\phi(\lambda_1)$ and $\phi(\lambda_2)$, respectively). This introduces an added phase difference term $\Phi(\Lambda)$ based on the two accrued phases (due to the object) in the argument of the cosine term of Eq.~\ref{g(2)_explicit}. With all of the above arguments, the second-order correlation of Eq. \ref{g(2)_explicit} modifies to: 
\begin{equation}\label{g_x_2}
g^{(2)}_x (0,\Delta t) = 1 - \frac{\left|g^{(1)}_{\lambda_1} \right|\left|g^{(1)}_{\lambda_2}\right|}{2} \cdot \cos(\Delta \omega \Delta t + \Phi(\Lambda))
\end{equation}
Comparing this to the intensity in Eq.\ref{int_at_port}, we observe that it has the same form as standard electric field phase interference, but with $\Delta \omega$ instead of $\omega$ representing the frequency of a synthetic (beat) wave. 
Stated differently, the $g^{(2)}$ interferometer for the synthetic wavelength is analogous to a conventional interferometer for an optical physical wave, with the only difference being a reduced amplitude of 0.5 in the interference term.  This therefore represents the direct interferometric detection of the synthetic fields using an intensity interferometer.\\
For wavelengths that are very close spectrally i.e. for small $\Delta\omega$, it is not necessary to measure the intensity at each wavelength separately (which can be technically challenging) but we can instead measure them simultaneously at single output port of an interferometer with a single camera. The $g^{(2)}$ auto-correlation equation then becomes (see SM for details)
\begin{equation}\label{eq:g2bucket}
\begin{aligned}
g^{(2)}_a (0,\Delta t) &= 1 + \frac{{\left|g^{(1)}_{\lambda 1}\right|}^2}{8} 
+ \frac{{\left|g^{(1)}_{\lambda 2}\right|}^2}{8}  \\
&\quad + \frac{\left|g^{(1)}_{\lambda 1}\right| \left|g^{(1)}_{\lambda 2}\right|}{4} 
\cdot \cos(\Delta \omega \Delta t + \Phi(\Lambda))
\end{aligned}
\end{equation}
This shows that measurement at one single output results in additional (constant) background terms (the auto-correlation of the two wavelengths) and a further reduction in the beating amplitude to 1/4 compared to standard holography.\\
The full synthetic field can then be extracted from the measured intensity correlations with the same phase retrieval techniques used for conventional holography. For our study, we employ the standard phase stepping technique  \cite{Yamaguchi:97} that allows to extract the synthetic hologram. For the phase stepping, we measure the time-averaged second-order correlations thrice, each time with a different value of $\Delta t$, such that the time delay introduces a phase difference of 0, $\pi/2$ and $\pi$, respectively. The synthetic hologram after subjecting the measurements to phase stepping analysis can be expressed as (see Methods): 
\begin{equation}\label{eq:g2synfield}
    \widetilde{E}(\Lambda) = \frac{\left|g^{(1)}_{\lambda_1}\right| \left|g^{(1)}_{\lambda_2}\right|}{4} e^{i \Phi(\Lambda)}
\end{equation}  
This is formally similar to Eq.~(\ref{eq:synfield}) obtained from standard phase holographic methods, but now measured using $g^{(2)}$ correlations. We point out that the phase stepping phase difference is not to be confused with the randomized phase $\Delta \varphi$ discussed earlier. They operate on separate temporal and length scales and have different implications. The former is due to the relatively larger optical delay (on the order of the synthetic wavelength $\Lambda$) introduced in one of the arms of the interferometer. It evolves over a longer timescale and is considered constant over all the elements of the ensemble. As such, it doesn't average out to zero for the operation defined as $\langle \cdot \rangle_t$ while calculating $g^{(2)}$ correlations. On the other hand, the phase $\varphi$ is random phase ascribed to each of the elements of the ensemble. These are generated by introducing a very small delay in one of the interferometer arm, on the order of the optical wavelength, and therefore, are on sub micron-scale. They change over the timescale corresponding to each acquisition of the ensemble. Hence, it varies in the range of $(0,2\pi)$ over all the copies of the ensemble (and so averages out to zero, as discussed for Eq.\ref{g(2)_explicit}). \\ 
Each acquisition in the ensemble must be captured with a camera exposure time shorter than the interferometer’s decorrelation time. This is essential to minimize noise, path
length jitter, and scatterer movement. However, this requirement in no way limits our proposed method because performing the ensemble average over hundreds of frames compensates for these short acquisition times and effectively enhances the signal-to-noise ratio. By increasing the number of acquired frames, we can mitigate the trade-off between exposure time and signal recovery, ensuring robust measurements. This is in stark contrast to traditional holography methods, which often require exposure times of milliseconds to seconds, which can be a limiting factor therein.\\
As discussed above, this technique is robust against phase noise owing to the time averaging requirement. This lends itself well to the case of imaging through dynamic scattering media. This is because as explained above phase randomization is necessary for the measurement of $g^{(2)}$ correlations. We can exploit the naturally induced phase noise from dynamic scattering to achieve the required random phases over different copies of the ensemble to carry out the averaging denoted by $\langle \cdot \rangle_t$ of Eq. (\ref{eq:g2generalintens}) which further simplifies the setup.

%
%
{\bf{Results: 2-Plane Scattering.}}
\begin{figure*}[t]
 \centering
 \includegraphics[width=0.95\textwidth]{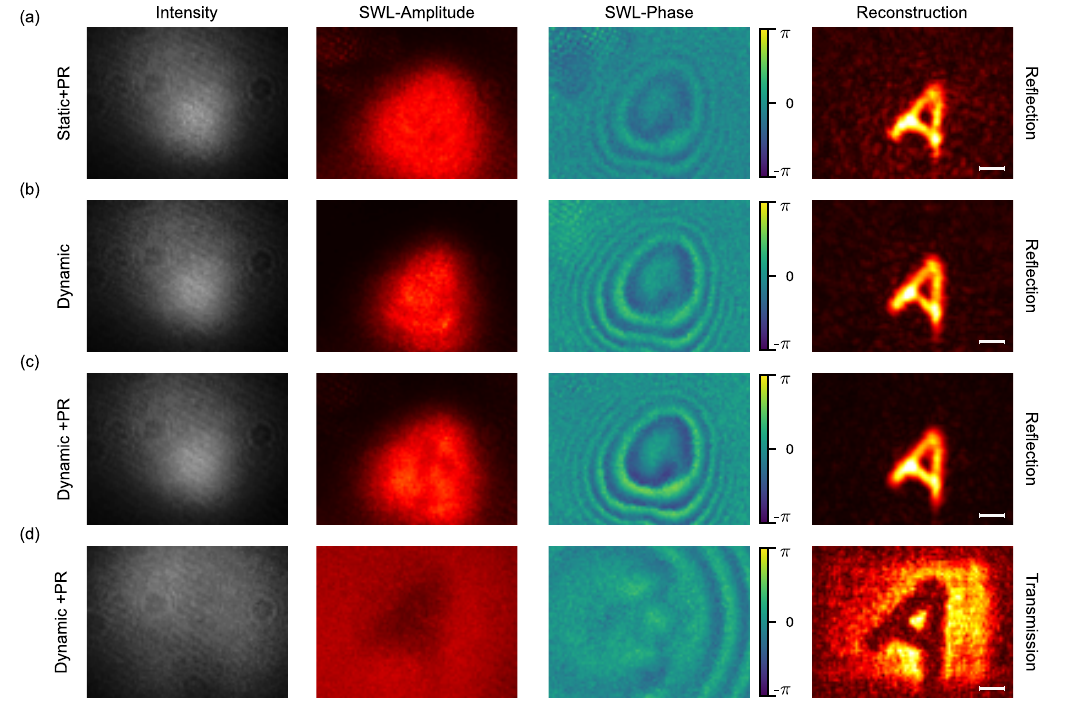}
\caption{\textbf{Recovering an object hidden by two scattering planes (ground glass diffusers):} The columns show (as indicated in the figure) the measured  intensity, synthetic wave (SW) amplitude and phase, and object reconstruction. (a), (b) and (c) are all performed in reflection:
(a) Static object with phase randomization (PR) applied in the reference path.  
(b) Same scene as (a), with random motion of one scatterer and no PR in the reference arm.  
(c) Dynamic scatterer with PR applied in the reference arm.  
(d) is performed in transmission for a dynamic scatterer with PR applied in the reference arm. Scale bar: 5 mm. The synthetic wavelength used for this was \(\Lambda = 0.68~\mathrm{mm}\)}
\label{fig:2planescat}
\end{figure*}
We begin by examining the simplest case of scattering at two distinct planes, implemented using optical diffusers in the experimental setup as shown in Figure~\ref{fig:experiment}. Here,  Scatterer A and Scatterer B are 220-grit and 120-grit optical diffusers, respectively. We investigate two measurement configurations: transmission, where the object is embedded between the two diffusers, and reflection, where the object is positioned behind a diffuser, with both illumination and imaging occurring from the same side, as illustrated in Figure~\ref{fig:experiment}.\\
In both configurations, the incident light undergoes an initial scattering at the first diffuser before interacting with the object. The light is subsequently scattered again by the second diffuser, leading to approximately two distinct plane of scattering. The scattered light is subsequently imaged with a camera, where it interferes with a reference beam. \\
A 3D-printed letter ``A'' was used as the object, as shown in Figure~\ref{fig:experiment}(c). The $ g^{(2)} $ holograms $\widetilde{E}(\Lambda)$ were generated using Eq.~\ref{eq:g2bucket} by capturing 1000 frames at three optical delay phase steps. The resulting holograms were then computationally backpropagated using the angular spectrum method (ASM) (see Methods) which reveals the object information. 

As discussed previously, phase randomization is necessary. Here, we investigate how the reconstruction changes when we apply phase randomization in two ways: first, by controlled phase randomization in the reference arm, as in standard \( g^{(2)} \), achieved by randomly moving the piezo shown in Figure~\ref{fig:experiment}. The second one, by utilizing the phase randomization from a dynamic scatterer.
Figure~\ref{fig:2planescat}(a)-(c) presents the results for the reflection configuration for different modes of static and dynamic scattering: \\ 
    \textbf{Static scatterer with reference phase randomization (Reflection)}: In Figure~\ref{fig:2planescat}(a), a static diffuser is used while applying phase randomisation (PR) to the reference arm, yielding a clear reconstruction of the letter ``A".\\ 
    \textbf{Dynamic scatterer without reference phase randomization (Reflection)}: In Figure~\ref{fig:2planescat}(b), phase randomization is induced solely by the dynamic scatterer, without additional phase randomization in the reference arm.\\
    \textbf{Dynamic scatterer with reference phase randomization (Reflection)}: Figure~\ref{fig:2planescat}(c) demonstrates that even when the scatterer is changing dynamically during measurement, the reconstruction remains intact, highlighting the method’s robustness to dynamic scattering.\\
    \textbf{Dynamic scatterer with reference phase randomization (Transmission)}: Additionally, we validated the method’s effectiveness in the transmission configuration. The results, shown in Figure~\ref{fig:2planescat}(d), present the case with a dynamic scatterer and reference phase randomization, yielding a clear reconstruction of the letter ``A". This confirms the robustness of the technique in both reflection and transmission geometries.\\
\begin{figure*}[t]
 \centering
 \includegraphics[width=0.95\textwidth]{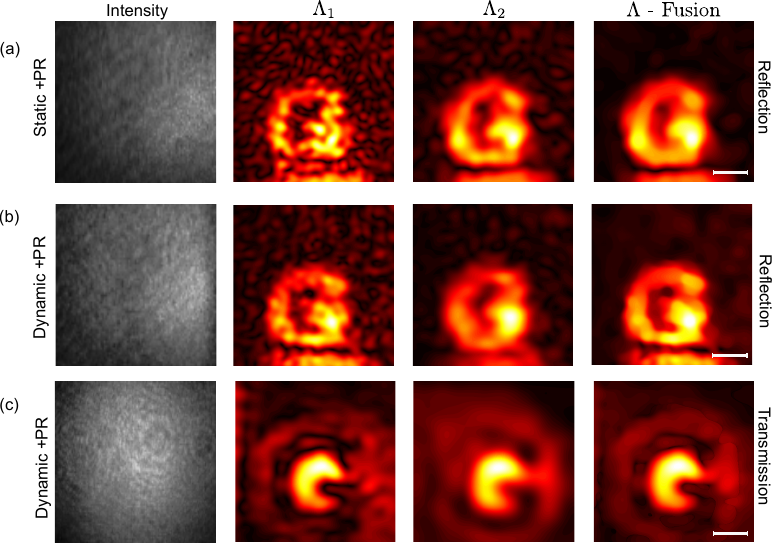}
 \caption{\textbf{Recovering an object hidden by volume scatterers:} The columns display intensity, reconstruction for each wavelength, and object reconstruction by fusing information from both, as indicated at the top of each column. 
Reflection($\Lambda_1= 2;\Lambda_2= 3$ mm):  
(a) Static object with phase randomization (PR) applied in the reference path.  
(b) Dynamic scatterer with PR applied in the reference arm.  
Transmission($\Lambda_1= 3 , \Lambda_2 = 5$ mm):  
(c) Dynamic scatterer with PR applied in the reference arm. Scale bar: 5 mm.}
\label{fig:bulkscat}
\end{figure*}
The synthetic wavelength $\Lambda$ used for these results was 0.68 mm which was generated using optical wavelengths of $854.31$ nm and $853.24$ nm. Comparing the different reconstructions, we not only confirm the ability to measure through dynamic scattering, but also observe a reduction in noise in the dynamic case compared to the static case. This improvement arises from the continuous variation of the speckle pattern in the dynamic case, effectively improving the sampling of the hologram. This can also be understood by noticing that in contrast, a static speckle pattern would lead to regions with zero intensity, providing no useful information at this point, whereas a dynamic change of this pattern leads to changes in the intensity at every pixel.
\\

\begin{figure}[t]
 \centering
 \includegraphics[width=0.5\textwidth]{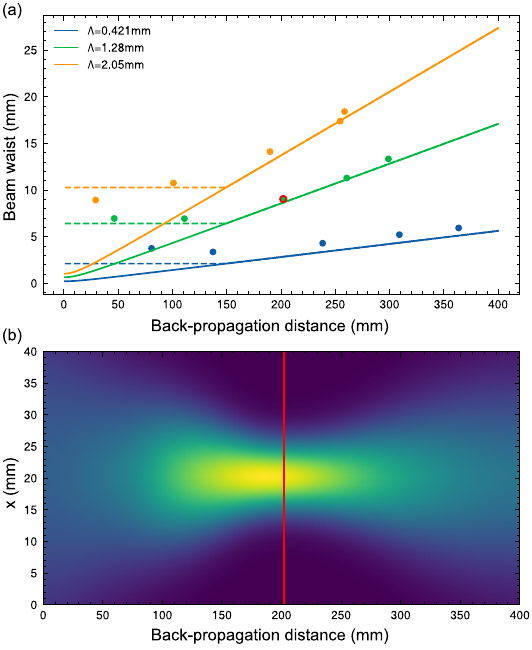}
\caption{{\textbf{Diffraction-limited resolution:} (a) Measured diffraction beam waist size (dots) for a a point-source (optical fiber tip) and the theoretical Abbe limit (solid line) as a function of the numerical back-propagation distance. The dashed line and plateau arises due to the limited NA of the fiber tip,  which constrains the system's numerical aperture (NA) from short distances up to approximately 150 mm. (b) An $x-z$ cutout of the back-propagated synthetic field for the fiber tip at 200 mm from the diffuser with $\Lambda = 1.28$ mm. The corresponding measurement point in (a) is highlighted with a red border.}
}
\label{fig:resolution}
\end{figure}

{\bf{Volume Scattering.}}
We then demonstrate more challenging cases, specifically for volume scatterers for both transmission and reflection configurations. 
For the reflection configuration, we employed a scatterer with a thickness of $2.7 \, \text{cm}$ and a reduced scattering coefficient, $\mu_s^* = 0.22 \, \text{mm}^{-1}$, corresponding to an average of 12 scattering events (see Supplementary for details). For the transmission configuration, we used two thinner yet stronger scattering elements to embed the object. The volume scatterer on the camera side was $1.5 \, \text{cm}$ thick with a reduced scattering coefficient of $\mu_s^* = 1.1 \, \text{mm}^{-1}$, which is comparable to the human tissue. The scatterer on the laser side was $0.8 \, \text{cm}$ thick with a reduced scattering coefficient of $\mu_s^* = 1.1 \, \text{mm}^{-1}$, leading to an average of 25 scattering events. \\
Given the increased scattering in these cases, the choice of $\Lambda$ becomes crucial. There is an inherent trade-off between resolution and signal-to-noise ratio (SNR) or reconstruction quality \cite{Willomitzer2021FastHolography}. Stronger scattering requires a larger $\Lambda$ in order to allow imaging at deeper depths (i.e. larger $\Lambda$ are less affected by scattering), which however, compromises resolution (i.e. the standard Rayleigh criterion applies and the smallest resolvable features scale with $\Lambda$).\\
However, image quality is then improved by deploying iterative optimization algorithms and total variation (TV) denoising techniques to converge to improved reconstructions of the object and also implement the fusion of images acquired at different wavelengths $\Lambda$ thus merging low-resolution but higher SNR images from longer synthetic wavelengths with higher-resolution but lower SNR images from shorter synthetic wavelengths  (see Supplementary Material for details).\\
In our experimental setup, we first investigated a static scatterer with a 3D-printed letter ``G" object, in reflection. We performed measurements with two synthetic wavelengths, $\Lambda = 3.08 \, \text{mm}$ and $\Lambda = 5.33 \, \text{mm}$. To generate these synthetic wavelengths the first optical wavelength was kept constant at $\lambda_1 = 854.31\, \text{nm}$, while the second was changed to $\lambda_2 = 854.07\, \text{nm}$ and $\lambda_2 = 854.17\, \text{nm}$, respectively. The corresponding reconstruction results are shown in Fig.~\ref{fig:bulkscat}(a), where we present the back-propagated fields for each wavelength and the computationally fused result. As expected, the smaller $\Lambda$ provides more details but exhibits more noise. The iterative algorithm aims to balance these effects, optimizing both SNR and resolution.\\
Next, we tested the robustness of our method to dynamic changes by randomly moving the volume scatterer. The result, shown in Fig.~\ref{fig:bulkscat}(b), demonstrates a good reconstruction that preserves the key structural features of the object. Notably, there is a slight gain in SNR, which, as mentioned, is attributed to the fact that the speckle pattern is constantly changing thus removing noise artefacts that might appear with static or slowly moving speckle.\\
Finally, we repeated the experiment in transmission with the object placed between two different volume scatterers, as described earlier. The results, shown in Fig.~\ref{fig:bulkscat}(c), further highlight the robustness and effectiveness of the approach.\\
{\bf{Image quality and diffraction limit.}}
As outlined in Ref.~\cite{Willomitzer2021FastHolography,willomitzer2024synthetic}, the Rayleigh quarter-wavelength criterion for wavefront aberrations \cite{rayleigh1879xxxi}
remains applicable to measurements conducted at the synthetic wavelength. According to this criterion, a clear synthetic wavelength hologram can be formed when the maximum path length variation $\Psi$ induced by the scatterer, in conjunction with geometric factors, is less than one-quarter of the wavelength $\Lambda$. Hence, if we wish to minimise image degradation, the synthetic wavelength $\Lambda$ has to be:
\begin{equation}\label{eq:rayleighcrit}
   \Lambda > 4\,\Psi
\end{equation}
For a given scattering problem, this will act as a soft lower limit for the required synthetic wavelength. However,  the synthetic wavelength also adheres to Abbe's diffraction limit \cite{abbe1873beitrage,Willomitzer2021FastHolography, willomitzer2024synthetic}:
\begin{equation}
    \Lambda = 2 \, \text{NA} \, \delta_{x,y}
\end{equation}
where $\delta_{x,y}$ is the resolution and NA is the numerical aperture.  
Hence, the selection of $\Lambda$ requires a compromise between signal quality and resolution. In practice, this means there is a trade-off between maximizing signal-to-noise ratio (SNR) and maintaining sufficient resolution based on the object to be imaged.\\
{Figure~\ref{fig:resolution}(a) shows measurements of the waists of a back-propagated diffraction spot at the synthetic wavelength \(\Lambda\), along with a comparison to theory (see Supplementary Material for details). The diffraction spots were generated by a point source (fiber-tip) hidden behind a 120-grit ground glass diffuser. These results clearly demonstrate that the synthetic wavelength is unaffected by the scatterer, achieving diffraction-limited resolution as if the scatterer were not present. This allows for selecting an optimal wavelength for the best SNR or fusing multiple wavelengths to balance SNR and resolution.}



\textbf{{Conclusion.}}
In this work, we introduced a novel method for generating synthetic wavelength holograms for imaging through scattering using $g^{(2)}$ correlations, circumventing the need for phase stability and effectively, leveraging the phase randomisation that is intrinsic to dynamically changing complex media. Our approach was demonstrated to effectively operate in challenging scenarios, including imaging through dynamic and thick diffusers. We highlighted the underlying technique by progressing through steps of increasing image-reconstruction difficulty, starting from thin diffusers in transmission or reflection mode, to a thick static diffuser, and finally to the dynamic thick diffuser.
%
%
Our findings highlight the importance of selecting an optimal synthetic wavelength to balance signal-to-noise ratio (SNR) and resolution, as well as the benefits of incorporating multiple synthetic wavelengths and computational techniques for improved image reconstruction. 
The intensity-correlation approach performs best when the scatterers are dynamically changing. This is particularly relevant in medical bioimaging settings. In future work, we aim to test this method for applications such as imaging through skin and other biological tissues.

\section*{Methods}
{\bf{Experimental Setup.}}
The experimental setup is based on a fiber-based Mach-Zehnder interferometer, as shown in Fig. \ref{fig:experiment}. A Toptica DFB Pro 855 nm laser with two separate laser heads are directly coupled into  a fiber beamsplitter (Thorlabs TW850R5A2) serving as the input beamsplitter. One output is collimated in free-space (Throlabs F810APC collimator), forming the reference arm. The collimator is mounted on a piezo stage for phase randomization and is in turn mounted on a courser but longer tuning range motorized translation stage for $\Lambda$ phase stepping. The other output is also collimated in free space and is directed onto the scattering  samples, either behind or in front ofthe object, depending on whether transmission or reflection measurements are performed.\\
The interferometer output, a 50:50 free-space beamsplitter (Thorlabs BS033), combines the reference and object beams, and a 50 mm lens images the scatterer’s surface plane onto the camera. To ensure the speckle size is not smaller than the camera pixels, an iris is placed infront of the imaging lens for control when necessary. The signal is captured by a Basler acA720-520um camera, which operates at 500 Hz with a resolution of 720×540 pixels. Additionally, the scatterer’s independent dynamic movement was generated using a conventional vibration motor.\\

{\bf{ $g^{(2)}$ Synthetic Wavelength Hologram.}}
For the SWL hologram generation at a single measurement point, we acquire 1000 frames to compute $g^{(2)}$, as defined in Eq.~\ref{eq:g2bucket}. Since our equation contains three unknowns—$\left|g^{(1)}_{\lambda_1}\right|$, $\left|g^{(1)}_{\lambda_2}\right|$, and $\Phi(\Lambda)$—only three phase steps are required for full field recovery via phase-stepping.\\
We can express $g^{(2)}_0(\Delta t)$ in the simplified form:
\begin{equation}
    g^{(2)}_0 (\Delta t) = a + b \cos(\Delta \omega \Delta t + c)
\end{equation}
where phase steps of $0, \pi/2, \pi$ correspond to optical delays of $0, \Lambda/4, \Lambda/2$, enabling field retrieval. The parameters $a$, $b$, and $c$ are obtained as:
\begin{equation}
    \begin{aligned}
        a &= \frac{g^{(2)}_0 (0) + g^{(2)}_0 (\pi)}{2}, \\
        b &= \sqrt{{g^{(2)}_0 (0)}^2 + {g^{(2)}_0 (\pi/2)}^2}, \\
        c &= \arctan \left( \frac{g^{(2)}_0 (0)-a}{g^{(2)}_0 (\pi/2)-a} \right).
    \end{aligned}
\end{equation}
Finally, the recovered field is given by:
\begin{equation}
    \widetilde{E}(\Lambda) = b e^{i c},
\end{equation}
which is equivalent to Eq. (\ref{eq:g2synfield}).\\

{\bf{Computational reconstruction.}}\label{sec:comprec}
The SWH is computationally backpropagated to retrieve the optical field in the plane of the object using the angular spectrum method. In this method, the synthetic field measured on the surface of the scatterer ($E_{m}(x,y,z)$) is modeled as the propagated field of the object plane field ($E(x,y,0)$), and the latter is numerically evaluated using the integral relationship: 
\begin{equation}\label{eq:integral}
\begin{aligned}
    E_{m}(x,y,z)= \int\int df_{x}df_{y}[\hat{E}(f_{x},f_{y},0)exp(iz\alpha)]&&\\exp[i2\pi(f_{x}x+f_{y}y)]
\end{aligned}
\end{equation} 
Here, $x,y$ are the transverse coordinates and $z$  the propagation distance.
$\hat{E}(f_{x},f_{y},0)$ is the Fourier transform of the source optical field $E_{0}(x,y,0)$. $f_{x}$ and $f_{y}$ are the spatial frequencies corresponding to the $x$ and $y$ axes and $\alpha$ is given by:
\begin{equation}\label{eq:alpha}
\begin{aligned}
\alpha = \sqrt{\left(\frac{2\pi}{\lambda}\right)^2-4\pi^2(f_{x}^2+f_{y}^2)}
\end{aligned}
\end{equation}
However, in challenging cases such as volume scattering, backpropagation alone may result in noisy images with speckle artifacts, reducing image quality.  
To mitigate this, we apply total variation (TV) regularized minimization to enhance reconstruction quality. Eq.(15) can be written in a compressed operator notation (see SM) as:
\begin{equation}\label{eq:Forw_model}
 E_{m} = \hat{A}E+\epsilon
\end{equation}
The operator $\hat{A}$ represents the integral transformation of Eq.(15) and $\epsilon$ represents measurement noise.The goal is to retrieve $E(x,y,0)$ by solving the optimization problem:
\begin{equation}\label{eq:cost_function_option2}
\begin{aligned}
    E^{o} =\underset{\widetilde{E}}{\arg\min}\;C(\widetilde{E}), \\
   C(\widetilde{E}) = \frac{1}{2}||\hat{A}\widetilde{E}-E_{m}||^2 +\alpha TV(\widetilde{E},\widetilde{E}^*)
\end{aligned}
\end{equation} 
 $E^{o}$ represents the optimal solution given by the algorithm which best approximates the field $E(x,y,0)$.The first term in the cost function enforces data fidelity, ensuring consistency with the measurements, while the TV regularizer, weighted by $\alpha$, suppresses noise and enhances reconstruction.  The cost function can be generalized to incorporate multiple measurements at different synthetic wavelengths:
 \begin{equation}\label{eq:cost_func2}
    C(\widetilde{E}) = \sum_{n=1}^{N}\frac{1}{2}||\hat{A}_{\lambda_n}\widetilde{E}-E_{m_n}||^2 + \alpha TV(\widetilde{E},\widetilde{E}^*),
\end{equation}
where $\lambda_n$ represents the synthetic wavelengths for the \(N\) measurements.
The value of the weight $\alpha$ was chosen empirically that gives the best reconstruction as done in standard optimization techniques.\\

\bigskip
\textbf{Funding}
K.K. acknowledges support from The SPIE Endowment Fund at the University of Glasgow. D.F. acknowledges financial support from the Royal Academy of Engineering Chairs in Emerging Technologies and the UK Engineering and Physical Sciences Research Council (projects no. EP/T00097X/1, EP/Y029097/1).\\
\textbf{Acknowledgments}
We thank Jack Radford and Enrique Pe\~na for their assistance with the phantoms.\\
\textbf{Author Contributions:}
K.K, D.F. and F.W. conceived and developed the idea. K.K and P.C. developed and built the experimental imaging setup. K.K and A.F. performed data analysis. A.F. developed the computational model. K.K and A.F. prepared the original draft. D.F., F.W., K.K, A.F, reviewed and edited the manuscript. D.F. and F.W. supervised the entire work.




\noindent The authors declare no conflicts of interest.

\textbf{Data availability:} Data underlying the results presented in this paper are available at DOI:xxx.


\newpage

%





\end{document}